\documentclass[aps,pre,twocolumn,showpacs,floatfix]{revtex4}
\usepackage{amsmath,bm,epsfig}
\usepackage{epsfig,color}
\usepackage{bm}

\renewcommand{\it}[1]{\textit{#1}}

\begin{document}
\setcounter{page}{1}

\title{Polymer Effects on Heat Transport in Laminar Boundary Layer Flow}
\author{Roberto Benzi$^1$, Emily S.C. Ching$^{2,3}$ and Vivien W.S. Chu$^2$}
\affiliation{$^1$Dip. di Fisica and INFN, Universit\`a ``Tor
Vergata", Via della Ricerca Scientifica 1, I-00133 Roma, Italy}
\affiliation{$^2$Department of Physics,
The Chinese University of Hong Kong, Shatin, Hong Kong}
\affiliation{$^3$Institute of Theoretical Physics, 
The Chinese University of Hong Kong, Shatin, Hong Kong}

\date{\today}

\begin{abstract}

{We consider a laminar Prandtl-Blasius boundary layer flow above a slightly
heated horizontal plate and study the effect of polymer additives on the heat
transport. We show that the action of the polymers can be understood
as a space-dependent effective viscosity that
first increases from the zero-shear value at the plate to a maximum 
then decreases exponentially
back to the zero-shear value
far away from the plate. We find that with such an effective
viscosity, both the horizontal and vertical velocities near the
plate are decreased thus leading to an increase in the friction drag and 
a decrease in the heat transport in the flow.}
\end{abstract}
%\pacs{47.27.te \hspace{0.5cm} Keywords: Polymer additives, heat transport,
%Blasius t convection}
\maketitle

It has been known for more than 60 years that
adding polymers into turbulent wall-bounded flows
can reduce the friction drag significantly
(see, for example, \cite{Sreeni,RMP} and references therein).
A reduction in friction drag is equivalent to an enhancement in mass
transport. Thus the effect of polymer additives on mass transport 
has been studied extensively during the past 60 or so years. On the other hand,
the effect of polymers additives on heat transport is much less studied.
Recently, an experimental study reported~\cite{AhlersPRL}
that adding polymers to turbulent Rayleigh-B{\'e}nard (RB) convection of
water, confined within a cylindrical cell heated from below and cooled
on top, reduces the heat transport.
In turbulent RB convection,
there is an exact balance~\cite{SiggiaReview} between the heat
transport and the energy and thermal dissipation rates. Contributions
to the energy
and thermal dissipation rates come from the bulk of the flow as well as
the boundary layers~\cite{GLtheory}.
For moderate thermal forcing as in the above-mentioned experimental study 
[using water at a moderate Rayleigh number (Ra) of $10^{10}$], 
contributions
from the boundary layers are dominant or significant~\cite{GLtheory}.
Furthermore, there has been experimental evidence~\cite{XiaJFM} that
the average velocity and temperature boundary layer profiles
in turbulent RB convection at moderate Ra could be described by
the profiles in steady-state Prandtl-Blasius boundary layer
flow~\cite{Landau,Schlichting}. These experimental observations 
thus suggest the
possibility of understanding the polymer effects on heat transport in
turbulent RB convection at moderate Ra by studying the effect of
polymer additives on heat transport in 
laminar Prandtl-Blasius boundary layer flow. 

We have carried out such a study. In this Letter, we report our work
and discuss our results. Physically, we can think of
this boundary layer flow as the flow near
the bottom plate of the convection cell in turbulent RB convection.

For the Prandtl-Blasius boundary layer flow, the equation of motion
for the velocity field $v_x \hat{x} + v_y \hat{y}$ is:
\begin{equation}
\label{PB}
v_x \partial_x v_x + v_y \partial_y v_x = \nu \partial^2_{yy} v_x
\end{equation}
Here $x$ denotes the direction along the plate, 
$y$ denotes the direction away from
the plate, and $\nu$ is the kinematic viscosity of the fluid.
The crucial point about Eq.~(\ref{PB}) is that the viscous term 
is balanced by the nonlinear advection term 
and that $\partial_y \gg \partial_x$. 
The latter is satisfied for flows with large
Reynolds number (Re).
Introducing the stream function 
\begin{equation}
\label{stream}
\Psi(x,y) \equiv  \sqrt{\nu x U} \phi(\xi) \qquad {\rm with} \qquad
\xi \equiv \sqrt{\frac{U}{\nu x}} y
\end{equation}
we obtain the famous Blasius equation 
\begin{equation}
2 \phi_{\xi \xi \xi } + \phi \phi_{\xi \xi} = 0
\label{BP-vel}
\end{equation} 
Here $\phi_{\xi}$ denotes $\partial_{\xi} \phi$. 
The boundary conditions are $v_x = v_y = 0$ at the plate and
$v_x \to U$ far away from the plate, leading to:
\begin{equation}
\label{velBC}
\phi(0) = \phi_{\xi} (0) = 0  \ ; \qquad \phi_{\xi}(\infty)=1
\end{equation}
Writing the temperature field as
\begin{equation}
T(x,y) = T_0 + (T_1-T_0)\theta(\xi)
\end{equation}
where $T_0$ and $T_1$ are respectively the 
temperatures at the plate and far away from the
plate (or at the center of the convection cell in the case of 
turbulent RB convection).
Then $\theta$ satisfies the equation
\begin{equation}
\label{temperature} 2 \theta_{\xi \xi} + {\rm Pr} \ \phi \theta_{\xi} = 0
\end{equation}
with the boundary conditions:
\begin{equation}
\label{tempBC}
\theta(0)=1  \ ; \qquad \theta(\infty)=0
\end{equation}
and Pr $= \nu / \kappa $ is the Prandtl number. 

Here, we want to investigate the effect of polymers on the heat transport
in this Prandtl-Blasius flow. The polymers produces an additional stress
in the momentum equation of the fluid.  
This polymer stress depends on the
amount of stretching of the polymers and is thus a function
of the dimensionless conformation tensor $R_{ij}$ of the polymers.
Let the vector $\vec{d}$ denote the polymer end-to-end distance
and $\rho_0$ be the polymer radius in the unstretched regime, then
$R_{ij}$ is the average over $N$ (where $N \gg 1$) 
polymers in a small region around the point $(x,y)$ 
of the product $d_i d_j/\rho_0^2$,
i.e., $R_{ij} = N^{-1} \Sigma d_i d_j/\rho_0^2$.
In the simplest Oldroyd-B model of polymers~\citep{Bird}, the polymer
stress is given by $ (\nu_p/\tau) ( R_{ij} - \delta_{ij} )$,
where $\tau$ is the (longest) relaxation time of the polymers, and
$\nu_p$ is the polymer contribution
to the viscosity of the solution at zero shear, which depends on
the concentration of the polymers.
Thus in the presence of the polymers,
the equation of motion for the velocity field is
modified by an additional stress that
depends on $R_{ij}$. By employing the same ideas
leading to the Prandtl-Blasius equation (\ref{PB}),
we have
\begin{equation}
\label{PBR} v_x \partial_x v_x + v_y \partial_y v_x = \nu
\partial^2_{yy} v_x + \frac{\nu_p}{\tau} \partial_y R_{xy}
\end{equation}

Applying the transformation by $\xi$ to Eq.~(\ref{PBR})
does not generally lead to a similarity solution in
that explicit appearance of $x$
remains in the equation for $\phi$. This is known in the literature.
Similarity solution have been obtained in some special cases with
the streamwise velocity and temperature going to some specific 
$x$-dependent functions when far away from
the plate~\cite{Olagunju2006,Bataller2008}. These velocity and
temperature boundary conditions do not, however, have direct physical 
relevance.
Here to circumvent this difficulty, we recall that the Prandtl-Blasius
flow is meant to be applicable when $x$ is large (such that $\partial_y
\gg \partial_x$). Thus we make the following approximations by
putting $x=L$ where $L \gg 1$ is the length of the plate:
\begin{eqnarray}
v_y &\approx& \frac{1}{2} \sqrt{\frac{\nu_0 U}{L}} (\xi \phi_\xi - \phi)
\label{approx1}\\
\partial_x &\approx& -\frac{\xi}{2L} \frac{d}{d\xi} \ ;
\qquad
\partial_y \approx \sqrt{\frac{U}{\nu_0 L}} \frac{d}{d\xi}
\label{approx2}
\end{eqnarray}
and we have also replaced $\nu$ in $\xi$ by
$\nu_0=\nu+\nu_p$, the zero-shear viscosity of the polymer-laden fluid.
That is, in the presence of polymers, we have
\begin{equation}
\xi = \sqrt{\frac{U}{\nu_0 x}} y \qquad {\rm with \ polymers}
\label{xinew}
\end{equation}
With these approximations,
the transformation of $\xi$ leads to a
similarity solution. The resulting
modified Prandtl-Blasius equation is:
\begin{equation}
\label{basic} -\frac{1}{2} \phi \phi_{\xi\xi} = (1-\gamma)
\phi_{\xi \xi \xi } + \frac{\gamma}{{\rm Wi} \sqrt{\rm Re}}
\frac{d}{d \xi} R_{xy}
\end{equation}
where the Weissenberg number (Wi) and the Reynolds number (Re) are
defined as
\begin{equation}
{\rm Wi}\equiv \frac{\tau U}{L} \ , \qquad
{\rm Re} \equiv \frac{UL}{\nu_0}
\end{equation}
and $\gamma \equiv \nu_p /\nu_0$
is a function of the polymer concentration.
As usual in the Prandtl-Blasius approximation, all terms of the order
of 1/Re are nelgected in Eq.~(\ref{basic}).

We are interested in studying whether and 
how the heat transport is affected by the polymers.
In turbulent RB convection, it is common to measure the heat flux $Q$
in terms of the
dimensionless Nusselt number (Nu), which is the ratio of $Q$ to the heat
flux when there is only conduction and is defined by
\begin{equation}
{\rm Nu}=  \displaystyle \frac{Q}{2k(T_1-T_0)/H} = \displaystyle
\frac{ \langle \displaystyle -\frac{\partial T}{\partial y}
\bigg |_{y=0} \rangle_{A} }{2(T_1-T_0)/H}
\end{equation}
Here $k$ is the thermal conductivity of the fluid,
$H$ is the height of the convection cell, and $\langle \ldots \rangle_A$
is the average over the cross section of the cell. For
the Prandtl-Blasius flow, taking $H=L$ and dropping the numerical
factor, Nu can be estimated as
\begin{equation}
{\rm Nu}= \sqrt{\frac{UL}{\nu_0}} \ [-\theta_{\xi}(0)]
\label{Nutheta}
\end{equation}

We want to state our major result: we will show that 
the action of the polymers is to give rise to a space-dependent 
effective viscosity that increases from the zero-shear value at the
boundary.
This increase of viscosity leads to an enhancement in drag, which in
turn induces a reduction in heat flux. To solve Eq.~(\ref{basic}),
one needs to supplement
it with specific information of $R_{xy}$. In a fluid flow of
velocity $\vec{v}$, the dimensionless
polymer end-to-end distance $\vec{l}=\vec{d}/\rho_0$ obeys the
differential equations
\begin{equation}
\label{polymer} \frac{d l_i}{dt } = - \frac{1}{2 \tau} (l_i
- l_{0i}) + l_j \partial_j v_i+ {\rm thermal \ noise}
\end{equation}
where $l_{0x}= \cos \alpha$, $l_{0y}= \sin \alpha$ and $\alpha$ 
is a random angle uniformly distributed in $[0,2\pi]$. 
For the two-dimensional Prandtl-Blasius
flow and neglecting thermal noise, we can rewrite Eq.~(\ref{polymer}) as
\begin{eqnarray}
\label{polx}
\frac{L}{U} \frac{d l_x}{d t} &=& - \frac{(l_x-l_{0x})}{2 {\rm Wi}}
- \frac{1}{2}\xi \phi_{\xi \xi} l_x + \sqrt{\rm Re} \phi_{\xi \xi} l_y \\
\frac{L}{U}\frac{d l_y}{d t} &=& - \frac{(l_y-l_{0y})}{2 {\rm Wi}}
- \frac{1}{4 \sqrt{\rm Re}} \xi^2 \phi_{\xi\xi} l_x   
+ \frac{1}{2} \xi \phi_{\xi \xi} l_y
\label{poly} 
\end{eqnarray}
In order to obtain $l_i$ as a function of $\xi$, we assume
that each polymer follows the streamline of the flow such that
$d\xi/dt \approx -(U/2L)\phi$.

The quantity $R_{xy} $ is the average of $l_x l_y$ over all the
polymers contained in a small volume centered near the point
$(x,y)$ and over the angle $\alpha$. 
Equations~(\ref{basic}), (\ref{polx}) and (\ref{poly})
are to be solved consistently. The procedure of how these
equations are solved is rather involved and the details 
will be discussed in a forthcoming paper~\cite{long}. 
Here we show that one can already obtain useful physical insights 
by assuming a balance, for any $\xi$, between 
the stretching of the polymers due to the velocity field
and the relaxation, represented by the term with the factor $1/{\rm Wi}$.
This is equivalent to say that, in the small-Wi limit,
we can neglect the LHS of Eqs.~(\ref{polx}) and (\ref{poly}).
Neglecting terms $O(1/\sqrt{\rm Re})$ and averaging over $\alpha$, 
we obtain, to the order in ${\rm Wi}^2$:
\begin{equation}
\label{result1}
R_{xy}  = 
(1+ \xi \phi_{\xi \xi} {\rm Wi}) \phi_{\xi \xi} {\rm Wi} \sqrt{\rm Re} 
\end{equation}
Substituting Eq.(\ref{result1}) in Eq.~(\ref{basic}),
we have
\begin{equation}
\label{result2}
2 \phi_{\xi \xi \xi } +  
\phi \phi_{\xi\xi} +
2 \partial_{\xi} [ f(\xi) \phi_{\xi \xi}] = 0
\end{equation}
where $f(\xi) \equiv  A \xi \phi_{\xi \xi}$
and $A = {\rm Wi} \gamma$
is a function of Wi and the polymer concentration. Equation~(\ref{result2})
demonstrates that the effect of the polymers is equivalent to
introducing a space dependent viscosity.
To see this, go back to Eq.~(\ref{PBR}) and
assume that $\nu_p R_{xy}/\tau$ is equivalent to the product of
a $y$-dependent effective viscosity, $\nu_p+\nu_0 f(y)$, and 
the velocity gradient, $\partial_y v_x$, we obtain:
\begin{equation}
\label{PBR2}
v_x \partial_x v_x + v_y \partial_y v_x = 
\nu \partial_{yy}^2 v_x +
\partial_y \{[\nu_p +\nu_0 f(y)] \partial_y v_x \} 
\end{equation}
which is exactly equivalent to Eq.~(\ref{result2})
once the rescaling in the variable $\xi$ is performed.

\begin{figure}[htb]
        \begin{center}
        \includegraphics[width=0.35\textwidth]{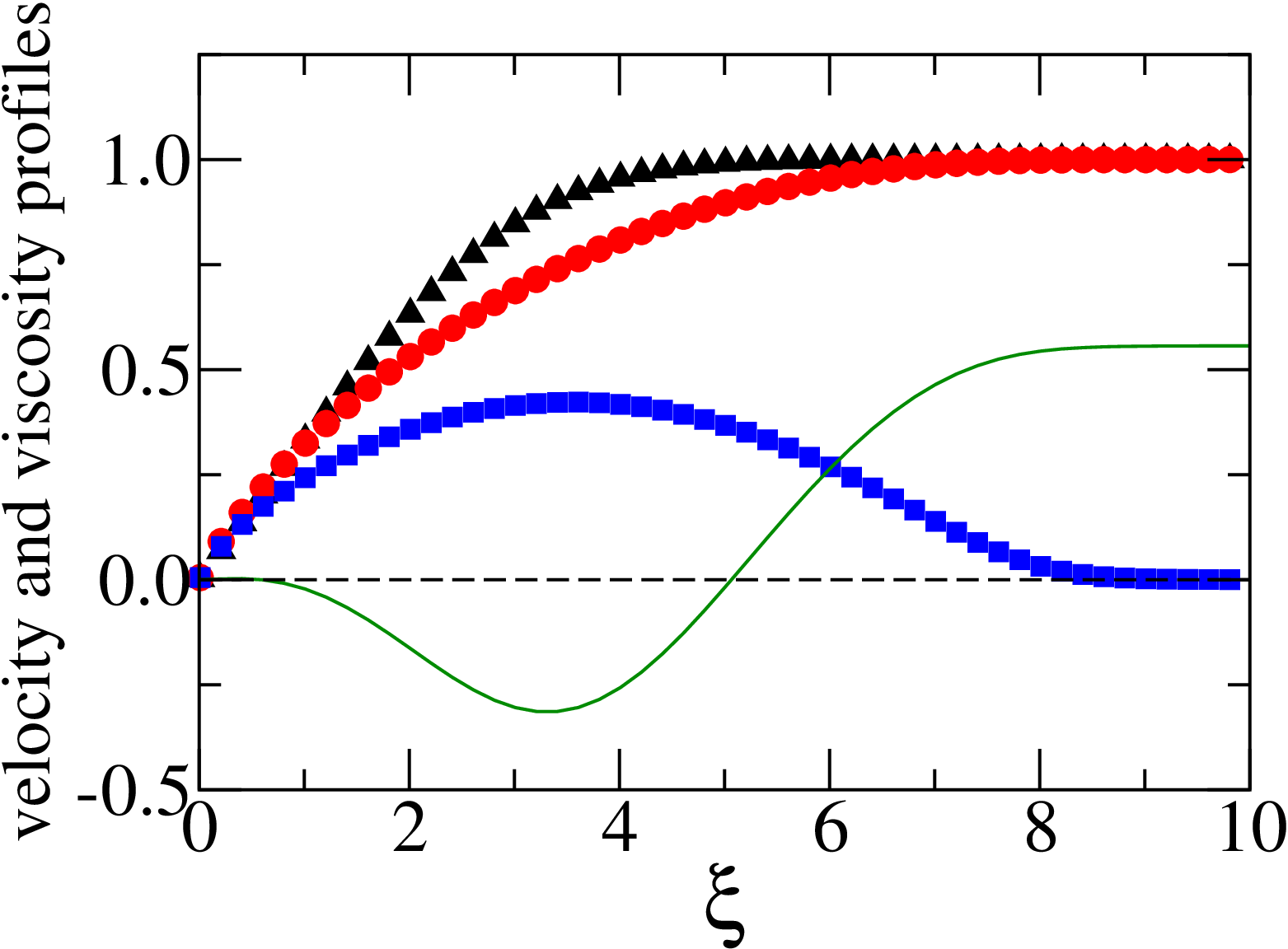}
        \end{center}
        \caption{Viscosity and velocity profiles 
obtained by solving Eq.~(\ref{result2}) with $A=0$ and $A=1$. 
We show the space-dependent viscosity $f(\xi)$ for $A=1$~(squares),
the horizontal velocity,
$v_x^{0}(\xi)/U$~(triangles) and $v_x^p(\xi)/U$~(circles), 
without ($A=0$) and with polymers ($A=1$), and
the difference in vertical velocity, 
$[v_y^{p}(\xi)-v_y^{0}(\xi)]/(U/2\sqrt{\rm Re})$~(solid
line).}
  \label{fig1}
\end{figure}

We solve Eq.~(\ref{result2}) for $A=1$, and show
$f(\xi)$ in Fig.~\ref{fig1}. It can be seen that
$f(\xi)$ increases from zero at the boundary and then
decreases almost exponentially towards zero. This increase
of the effective viscosity near the plate allows us to predict
the effect of the polymer on the flow. 
First, we expect an increase in the effective viscosity
near the plate would give rise to an increase of the drag.
We measure the drag by the drag coefficient:
\begin{equation}
C = \frac{\nu_0 \partial_y v_x \big|_{y=0}}{U^2/2} =
\frac{1}{2 \sqrt{\rm Re}} \
\phi_{\xi \xi}(0)
\label{drag}
\end{equation}
and compare $C$ to the value $C_0$ for the flow without polymers.
We expect $C/C_0$ would be greater than $1$ for all $A > 0$.
Such an enhancement in drag would in turn result in a reduction 
of the mass throughput in the horizontal direction and thus a decrease
in the horizontal velocity near the boundary.
These effects are indeed observed. In Fig.~\ref{fig1}
we show the horizontal velocity $v_x^0$ and $v_x^p$ 
respectively for the case without polymers
(denoted by the superscript $0$)
and with polymers (denoted by the superscript $p$), and
also compare the vertical velocities, $v_y^p$ and $v_y^0$, 
with and without polymers. It can be seen that in the presence of polymers, 
both the horizontal and vertical velocities 
decrease in the region near the plate.

Next, we show that drag enhancement implies a reduction of the 
heat transport or Nu.
Upon double integration of Eq.~(\ref{temperature}) by $\xi$, we obtain:
\begin{equation}
\label{nusselt}
- \theta_\xi(0) = 
\left\{\int_{0}^{\infty}d s_1  
\exp [- {\rm Pr} \ \int_{0}^{s_1} \phi(s_2) ds_2/2 ] \right\}^{-1}
\end{equation}
which tells us that Nu is a functional of $\Phi$, where
$\Phi(\xi) \equiv \int_0^{\xi} \phi(s) ds$. We can thus
calculate $\delta {\rm Nu}$, the variation in Nu induced by 
a variation in $\Phi$, $\delta \Phi(\xi)$:
\begin{equation}
\label{carina}
\delta {\rm Nu}  = \frac{{\rm Pr} \ {\rm Nu}^2}{2} 
\int_0^{\infty} \exp\left[-{\rm Pr} \ \Phi(s)/2 \right] \ \delta \Phi(s) \ ds
\end{equation}
Since $v_x = U \phi_{\xi}$, the mass throughput in the $x$ direction 
across a distance $\xi$ is given by $U \phi(\xi)$,
and thus a reduction in mass throughput implies an $\delta \Phi < 0$.
Therefore, Eq.~(\ref{carina}) 
shows that drag enhancement implies a reduction in Nu or the heat transport. 

\begin{figure}[b]
\vspace{0.5cm}
        \begin{center}
        \includegraphics[width=0.35\textwidth]{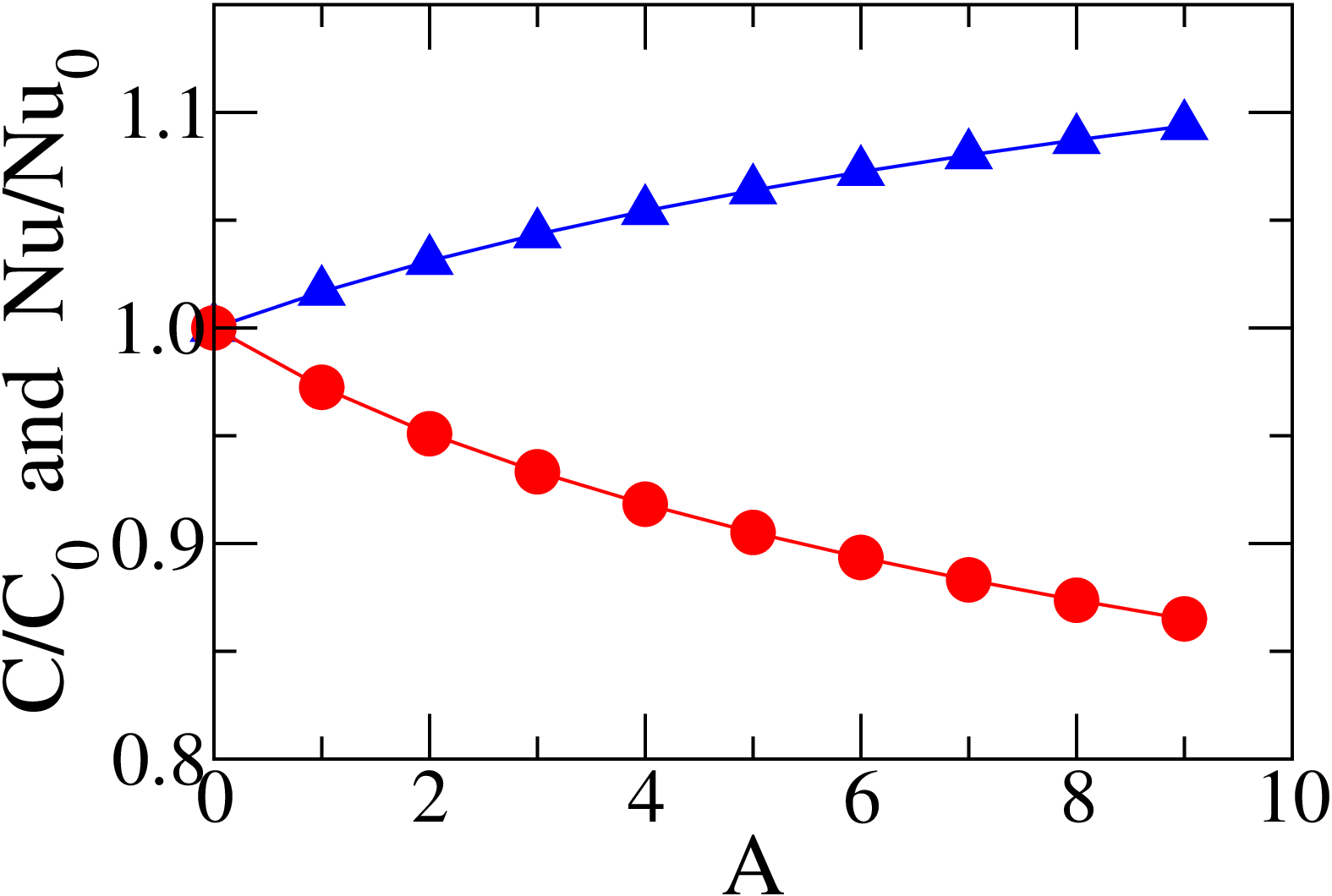}
        \end{center}
        \caption{Dependence of
$C/C_0$, (triangles)
and Nu/Nu$_0$, (circles)
on $A$ for $f(\xi)$ given by Eq.~(\ref{fdixi}).}
        \label{fig3}
\end{figure}

The above considerations are not restricted to the particular form 
of $f(\xi)=A \xi \phi_{\xi \xi}$ obtained
in the small-Wi limit but are true for a generic form of 
$f(\xi)$ which displays the same features of first increasing near the
boundary to some maximum then decreasing to zero far away from 
the boundary as shown in Fig.~\ref{fig1}. 
In particular, we show that this is the case 
by using a different form of $f(\xi)$:
\begin{equation}
\label{fdixi}
f(\xi) = A \xi \exp(-\xi)
\end{equation}
that has similar qualitative features.
The constant $A$ relates the polymer characteristics 
to the conformation tensor $R_{ij}$,
and generally increases with Wi and the concentration of the polymers.

\begin{figure}[htb]
\vspace{0.5cm}
        \begin{center}
\includegraphics[width=0.35\textwidth]{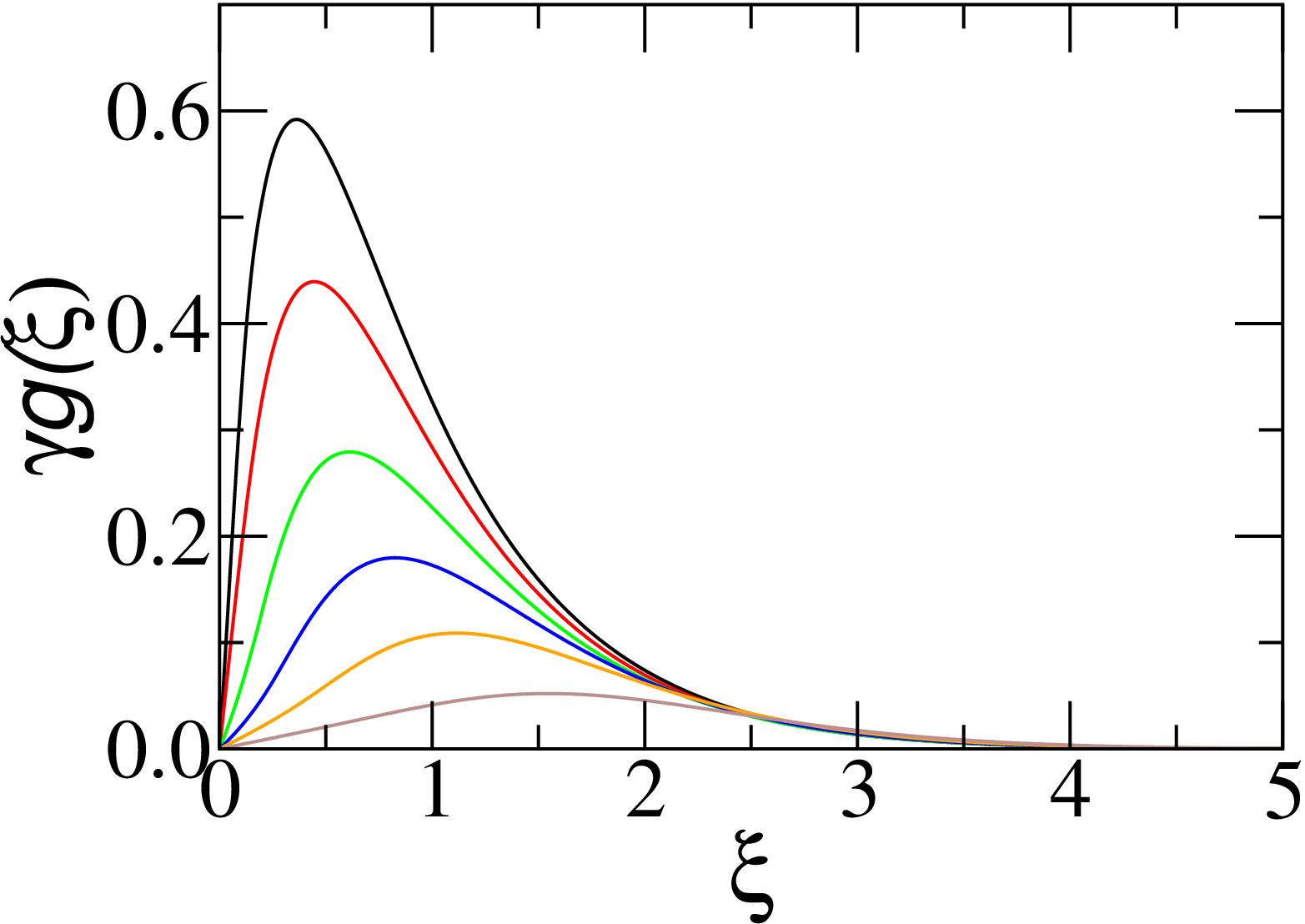}
        \end{center}
        \caption{$\gamma g(\xi)$ as a
function of Wi at fixed $\gamma=0.2$
obtained by the numerical solution of the full nonlinear problem,
Eqs.~(\ref{basic}), (\ref{polx}) and (\ref{poly}). Wi increases
from bottom to top, taking the values
0.5, 1.0, 1.5, 2.0, 2.5, and 2.8 respectively.}
        \label{fig2}
\end{figure}

Solving Eq.~(\ref{fdixi}) with Eq.~(\ref{result2}), 
we again find an increase in drag and a reduction in heat transport.
Moreover as shown in Fig.~\ref{fig3}, 
the extent of the effect
is increased by increasing $A$, i.e.,
by increasing the effect of the polymers
into the system. For the range of $A$ shown in Fig.~\ref{fig3},
the amount of the reduction of heat flux is quantitatively 
consistent with that observed in Ref.~\cite{AhlersPRL}.

\begin{figure}[h]

\vspace{1.0cm}
        \begin{center}
                \includegraphics[width=0.40\textwidth]{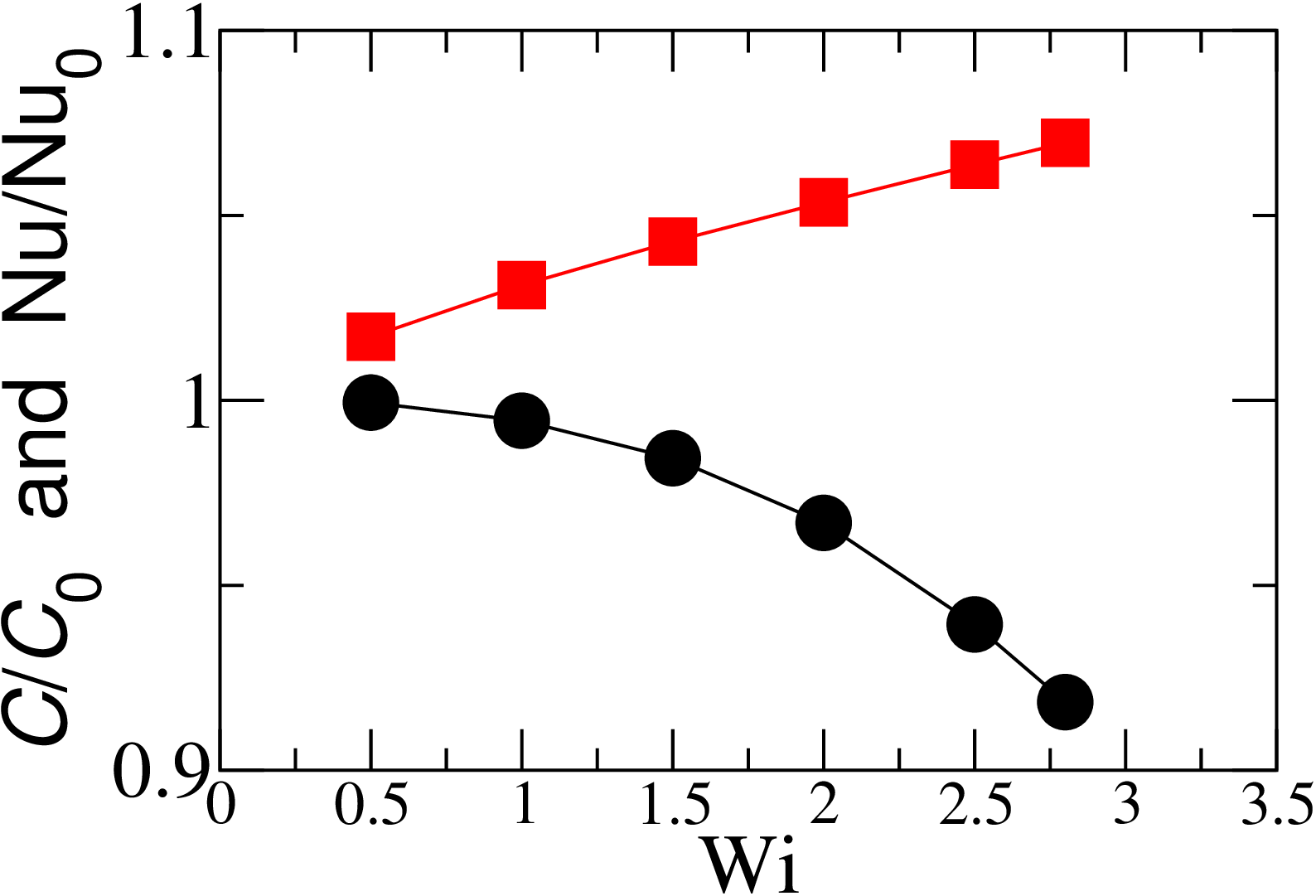}
        \end{center}
        \caption{Drag enhancement and heat transport reduction
for the full nonlinear problem and the
dependence of $C/C_0$~(squares) and
Nu/Nu$_0$~(circles) on Wi.}
        \label{fig4}
\end{figure}

In the remaining of this Letter, we show that  
the results that we have discussed 
capture the qualitative physical picture of the full nonlinear problem,
defined by Eqs.~(\ref{basic}), (\ref{polx}) and (\ref{poly}).
We have solved the full problem,
by integrating numerically the 
equations using an iterated procedure~\cite{long}.
In particular, we show that
$ R_{xy} = [1+g(\xi)] \phi_{\xi \xi} {\rm Wi} \sqrt{\rm Re}$ for some
function $g(\xi)$.
Substituting this into Eq.~(\ref{basic}) gives
\begin{equation}
\label{full2}
2 \phi_{\xi \xi \xi } + \phi \phi_{\xi\xi} + 
2 \partial_{\xi} [ \gamma g(\xi) \phi_{\xi \xi}] = 0
\end{equation}
Comparing Eq.~(\ref{full2}) 
with Eq.~(\ref{result2}) for the small-Wi limit, 
we see that $\gamma g(\xi)]$ plays 
the role of $f(\xi)$.
In Fig.~\ref{fig2}, we show $\gamma g(\xi)$ for different values Wi at a fixed
value of $\gamma=0.2$. Notice that the functional shape of $g(\xi)$ is similar
to the one used in Eq.~(\ref{fdixi}). Not surprisingly,  
we find drag enhancement and heat transport reduction for the full nonlinear
problem as shown in Fig.~\ref{fig4}. 

In summary, we have shown the following results:
\vspace{-0.3cm}
\begin{itemize}
\item[(1)] It is possible to write a self consistent set of 
equations for Prandtl-Blasius boundary layer flow with polymers. 
\vspace{-0.3cm}
\item[(2)] The physical effect of the polymers is equivalent to
introducing a space-dependent effective viscosity that is increasing 
near the boundary. 
\vspace{-0.3cm}
\item[(3)] An increasing viscosity near the boundary leads to 
drag enhancement, which in turn induces a reduction in heat transport.
\vspace{-0.3cm}
\end{itemize}

Our theory may thus explain the recent experimental results observed 
in turbulent RB convection~\cite{AhlersPRL}.

\bigskip

{\bf Acknowledgments}
This work was supported in part by the Hong Kong Research Grants Council
(CUHK 400708).


\begin{thebibliography}{99}
\bibitem{Sreeni} K.R. Sreenivasan and C. White, J. Fluid Mech. {\bf 409}, 149 (2000). 
\bibitem{RMP} I. Procaccia, V.S. L'vov, and R. Benzi, Rev. Mod.
Phys. {\bf 80}, 225 (2008).
\bibitem{AhlersPRL} G. Ahlers and A. Nikolaenko, Phys. Rev. Lett. 
{\bf 104}, 034503 (2010).
\bibitem{SiggiaReview} E.D. Siggia, Ann. Rev. Fluid
Mech. {\bf 26}, 137 (1994).
\bibitem{GLtheory} S. Grossmann and D. Lohse, J. Fluid Mech. {\bf 407}, 27 (2000).
\bibitem{XiaJFM} Q. Zhou, R.J.A.M. Stevens, K. Sugiyama, S. Grossmann,
D. Lohse, and K.-Q. Xia,
J. Fluid Mech. {\bf 664}, 297 (2010).
\bibitem{Landau} See, e.g., L.D. Landau and E.M. Lifshitz, \it{Fluid
Mechanics} (Pergamon Press, Oxford, 1987).
\bibitem{Schlichting} H. Schlichting and K. Gersten,
\it{Boundary-Layer Theory} (Springer, 8th ed. 2004).
\bibitem{Bird} R.B. Bird, O. Hassager, R.C. Armstrong, and C.F. Curtis,
\it{Dynamics of Polymeric Liquids} (Wiley-Interscinec 1987).
\bibitem{Olagunju2006} D.O. Olagunju, App. Math. Lett. {\bf 19},
432 (2006).
\bibitem{Bataller2008} R.C. Bataller, Phys. Lett. A {\bf 372}, 2431
(2008).
\bibitem{long} R. Benzi, E.S.C. Ching, and V.W.S. Chu, ``Heat transport
by laminar boundary layer flow with polymers", arXiv:1104.4526.
\end{thebibliography}
\end{document}